\documentclass[twocolumn,showpacs,preprintnumbers,amsmath,amssymb]{revtex4-1}
\usepackage{graphicx}
\usepackage{rotating}
\begin{document}

\renewcommand{\theequation}{\thesection.\arabic{equation}}

\newcommand{\re}{\mathop{\mathrm{Re}}}

\newcommand{\be}{\begin{equation}}
\newcommand{\ee}{\end{equation}}
\newcommand{\bea}{\begin{eqnarray}}
\newcommand{\eea}{\end{eqnarray}}


\title{A statefinder luminosity distance formula in varying speed of light cosmology}

\author{Adam Balcerzak}
\email{abalcerz@wmf.univ.szczecin.pl}
\affiliation{\it Institute of Physics, University of Szczecin, Wielkopolska 15, 70-451 Szczecin, Poland}
\affiliation{\it Copernicus Center for Interdisciplinary Studies, S{\l }awkowska 17, 31-016 Krak\'ow, Poland}

\author{Mariusz P. D\c{a}browski}
\email{mpdabfz@wmf.univ.szczecin.pl}
\affiliation{\it Institute of Physics, University of Szczecin, Wielkopolska 15, 70-451 Szczecin, Poland}
\affiliation{\it Copernicus Center for Interdisciplinary Studies,
S{\l }awkowska 17, 31-016 Krak\'ow, Poland}

\date{\today}

\input epsf

\begin{abstract}
We derive a luminosity distance formula for the varying speed of light (VSL) theory which involves higher order characteristics of expansion such as jerk, snap and lerk which can test the impact of varying $c$ onto the evolution of the universe. We show that the effect of varying $c$ is possible to be isolated due to the relations connecting observational parameters already by measuring the second-order term in redshift $z$ unless there is a redundancy between the curvature and an exotic fluid of cosmic strings scaling the same way as the curvature.
\end{abstract}

\pacs{98.80.-k; 98.80.Es; 98.80.Cq}

\maketitle

\section{Introduction}
\label{intro}
\setcounter{equation}{0}

The early idea of variation of physical constants \cite{varconst} has been established widely in physics both theoretically and experimentally \cite{uzan}. The gravitational constant $G$, the charge of electron $e$, the velocity of light $c$, the proton to electron mass ratio $\mu = m_p/m_e$, and the fine structure constant $\alpha$ may vary in space and time \cite{barrowbook}. The earliest and best-known framework for varying $G$ theories has been Brans-Dicke theory \cite{bd}. Nowadays, the most popular theories which admit physical constants variation are the varying fine structure constant $\alpha$ theories \cite{alpha}, and the varying speed of light $c$ theories \cite{VSL,uzanLR}. The latter, which will be the interest of our paper, allow the solution of the standard cosmological problems such as the horizon problem, the flatness problem, the $\Lambda-$problem, and has recently been proposed to solve the singularity problem \cite{JCAP13}. In this work we study the luminosity distance cosmological test for the varying speed of light theories.

Our paper is organized as follows. In Sec. \ref{VSL} we formulate the basics of the varying speed of light (VSL) theory and define
observational parameters such as the dimensionless energy density parameters $\Omega$, the Hubble parameter $H$, the deceleration parameter $q$, as well as the higher-order derivative parameters like the jerk $j$, the snap $s$ etc. \cite{jerk,snap,weinberg}, which may serve as indicators of the equation of state (statefinders)
and the curvature of the universe. In Sec. \ref{lumdist} we generalize the luminosity distance formula of Ref. \cite{BM99} to fifth order in redshift $z$ by the application of statefinders. In Sec. \ref{discussion} we discuss observational perspective to isolate variability of $c$ from the luminosity distance cosmological test.

\section{Varying speed of light theory and statefinders}
\setcounter{equation}{0}
\label{VSL}

Following Ref. \cite{VSL}, we consider the Friedmann universes within the framework of varying speed of light theories (VSL) described the metric
\be
ds^2 = - (dx^0)^2 + a^2(t) \left[ \frac{dr^2}{1-Kr^2} + r^2 (d\theta^2 + \sin^2{\theta} d\varphi^2) \right]~~,
\ee
where $dx^0 = c(t) dt$, for which the field equations read
\bea
\label{rho}
\frac{\dot{a}^2}{a^2} &=& \frac{8\pi G \varrho}{3} - \frac{Kc^2(t)}{a^2}~,\\
\label{p}
\frac{\ddot{a}}{a} &=& - \frac{4\pi G}{3} \left[ \varrho + \frac{3p}{c^2(t)} \right]~,
\eea
and the energy-momentum conservation law is
\be
\label{conser}
\dot{\varrho}(t) + 3 \frac{\dot{a}}{a} \left(\varrho(t) + \frac{p(t)}{c^2(t)} \right) = 3 \frac{Kc(t)\dot{c}(t)}{4\pi Ga^2}~.
\ee
Here $a \equiv a(t)$ is the scale factor, the dot means the derivative with respect to time $t$, $G$ is the gravitational constant, $c=c(t)$ is time-varying speed of light, and the curvature index $K=0, \pm 1$. In this paper we will use the following ansatz for the speed of light time evolution
\be
\label{c(t)}
c(t) = c_0 \left( \frac{a(t)}{a_0} \right)^n~~,
\ee
where $a_0$ is the current value (for $t=t_0$) of the scale factor and $n$ is a dimensionless parameter. The ansatz (\ref{c(t)}) differs from the one used in Ref. \cite{BM99} ($c(t) = c_0 a^n(t)$) since it interprets $c_0$ as the current value of the velocity of light (no matter the value of $n$) and allows $n$ to be dimensionless. If the speed of light does not vary during the evolution of the universe then there is a limit of the formula (\ref{c(t)}) which allows the constant speed of light which is $n \to 0$ giving $c(t) \to c_0$. Here also we have $\dot{c}/c = n \dot{a}/a$, so the speed of light grows in time for $n>0$, and diminishes for $n<0$.

The cosmological observables which characterize the kinematic evolution of the universe are \cite{PLB05}:
\\
\noindent the Hubble parameter
\be
\label{hubb}
H = \frac{\dot{a}}{a}~,
\ee
the deceleration parameter
\be
\label{dec}
q  =  - \frac{1}{H^2} \frac{\ddot{a}}{a} = - \frac{\ddot{a}a}{\dot{a}^2}~,
\ee
the jerk parameter \cite{jerk}
\be
\label{jerk}
j = \frac{1}{H^3} \frac{\dddot{a}}{a} =
\frac{\dddot{a}a^2}{\dot{a}^3}~,
\ee
and the snap \cite{snap}
\be
\label{snap}
s = -\frac{1}{H^4} \frac{\ddddot{a}}{a} =
-\frac{\ddddot{a} a^3}{\dot{a}^4}~.
\ee
We can carry on with these and define even the higher derivative parameters
such as lerk $l$ (crack), merk $m$ (pop), etc. \cite{PLB05,ANN06,gibbons} by
\be
\label{lerk}
x^{(i)} = (-1)^{i+1} \frac{1}{H^{i}} \frac{a^{(i)}}{a} = (-1)^{i+1} \frac{a^{(i)} a^{i-1}}{\dot{a}^{i}}~~,
\ee
where $i = 2, 3, ...$, and $a^{(i)}$ means the i-th derivative with respect to time while $a^{i}$ means the n-th power. We have consecutively:
$q$ for $i=2$, $j$ for $i=3$ etc. It is interesting that under the chosen ansatz (\ref{c(t)}) the variability of the speed of light enters already in the definition of the Hubble parameter since
\be
\label{hubbc}
H = \frac{\dot{a}}{a}~ = \frac{1}{n} \frac{\dot{c}}{c} \equiv \frac{1}{n} H_c,
\ee
where we have introduced the Hubble parameter for varying $c$ - $H_c$. In fact $H$ as given in (\ref{hubbc}) has a peculiar limit $n \to 0$
which is accompanied with $\dot{c} \to 0$.

A comparison of cosmological models with observational data requires the introduction of dimensionless density parameters \cite{AJIII}
\bea
\label{Omegadef1}
\Omega_{m0}  &=&  \frac{8\pi G}{3H_0^2} \varrho_{m0} ,\\
\label{Omegadef2}
\Omega_{K0}  &=&  \frac{Kc_0^2}{H_0^2a_0^2} ,\\
\label{Omegadef3}
\Omega_{\Lambda_0}  &=&  \frac{\Lambda_0 c_0^2}{3H_0^2} ,
\eea
for dust, curvature, and dark energy, respectively. The index "0" means that we take these parameters at the present
moment of the evolution $t=t_0$. In order to introduce these parameters we assume that the matter content of the universe is dust fulfilling the equation of state
\be
p_m = 0~~
\ee
and the conservation law
\be
\varrho_m a^3 = \frac{3}{8\pi G} C_m
\ee
where $C_m=$const., as well as evolving independently cosmological term with the pressure
\be
\label{pelam}
p_{\Lambda} = - \varrho_{\Lambda} c^2(t)~~,
\ee
and its mass density
\be
\label{rholam}
\varrho_{\Lambda} = \frac{\Lambda c^2(t)}{8\pi G}
\ee

Using Friedmann equation (\ref{rho}) and its time derivative and using the definitions of deceleration parameter (\ref{dec}) as well as omega parameters (\ref{Omegadef1})-(\ref{Omegadef3}) one obtains the relations
\be
\label{0th}
\Omega_{m0} - \Omega_{K0} + \Omega_{\Lambda 0} = 1~~,
\ee
and
\be
\label{1st}
\Omega_{\Lambda 0} (1+n) = \frac{1}{2} \Omega_{m0} - q_0 + n \Omega_{K0}~~,
\ee
which for $n=0$ generalizes the standard relation between dust matter and dark energy $\Omega_{\Lambda} = \Omega_{m0}/2 - q_0$. It tells us that in VSL theory due to the time-dependence of the curvature term in the Friedmann equation through $c^2(t)$, curvature already enters the observational quantities in the second order of the time derivative expansion. This can also be seen if one combines the relations (\ref{0th}) and (\ref{1st}) together i.e.
\be
\label{11st}
\Omega_{K0} = \frac{3}{2} \Omega_{m0} - (q_0 + n) - 1~~.
\ee
Taking the third derivative of the Friedmann equation (\ref{rho}) and using the definition of jerk (\ref{jerk}) one has
\be
\label{2nd}
j_0 = \Omega_{m0} + \Omega_{\Lambda 0}(n+1) - n \Omega_{K0}~~,
\ee
which by using (\ref{1st}) turns into
\be
j_0 = \frac{3}{2} \Omega_{m0} - q_0~~.
\ee
Fourth derivative of (\ref{rho}) and the definition of snap (\ref{snap}) gives next relation
\be
\label{3rd}
s_0 = 3\Omega_{m0} + n \Omega_{K0} - (n+1) \Omega_{\Lambda 0} + q_0 j_0~~,
\ee
which by the application of (\ref{1st}) can be expressed as
\be
\label{4th}
s_0 = \frac{5}{2} \Omega_{m0} + q_0(j_0 + 1) = 4 \Omega_{m0} + j_0 (q_0 - 1).
\ee
Note that if the curvature of the universe was neglected $\Omega_{K0} \approx 0$, then one would still have the VSL impact onto
(\ref{1st}), (\ref{11st}), (\ref{2nd}), and (\ref{3rd}) because of the dynamical involvement of $c=c(t)$ in the $\Lambda-$term (\ref{rholam}) so that VSL impact cannot be removed simply by disregarding the curvature.

There is yet another constraint for the observational parameters which comes from the 2nd law of thermodynamics and can be called ``entropic'' condition. Following Ref. \cite{youm} it reads as
\be
\label{entropy}
\dot{S} = \left(2 \varrho_m - \frac{\Lambda c^2(t)}{4\pi G} + 3 \frac{Kc^2(t)}{4\pi Ga^2} \right) \frac{c(t)\dot{c}(t) a^3}{T} \geq 0~~,
\ee
where $S$ - the entropy, $T$ - the temperature, and in our notation which uses (\ref{Omegadef1})-(\ref{Omegadef3}) it reads as
\be
\label{entropy1}
\Omega_{m0} - \Omega_{\Lambda 0} + \Omega_{K0} \leq 0~~
\ee
provided $\dot{c} < 0$ ($c$ diminishes) or the sign reverses if $\dot{c} > 0$.

In the standard limit $n \to 0$ the above relations read as \cite{ANN06}
\bea
\label{Omegarel1}
1 &=& \Omega_{m0} + \Omega_{\Lambda 0} + \Omega_{K0},\\
\label{Omegarel2}
\Omega_{K0} &=& \frac{3}{2} \Omega_{m0} - q_0 - 1,\\
\label{Omegarel3}
q_0 &=& \frac{1}{2} \left(\Omega_{m0} - 2\Omega_{\Lambda 0} \right) ,\\
\label{Omegarel4}
j_0 &=& \Omega_{m0} + \Omega_{\Lambda 0} ,\\
\label{Omegarel5}
s_0 &=& q_0 j_0 + 3 \Omega_{m0} ,\\
\label{Omegarel6}
l_0 &=& j_0^2 + 3 \Omega_{m0} (4 - 3 q_0).
\eea

\section{Statefinder redshift-magnitude formula in VSL theory}
\setcounter{equation}{0}
\label{lumdist}

In this Section we will follow the derivation of the VSL luminosity distance formula in Ref. \cite{BM99} which is based on standard calculation \cite{weinberg}, but we include jerk (\ref{jerk}), snap (\ref{snap}), and lerk (\ref{lerk}). The method of derivation was given in Ref. \cite{ANN06} and we will use it below. The scale factor $a(t)$ at any moment of time $t$ can be obtained as series expansion around $t_0$ as ($a(t_0) \equiv a_0$)
\bea
\label{seriesa}
&&a(t) = a_0 \left\{ 1 + H_0 (t-t_0) - \frac{1}{2!}q_0 H_0^2
(t-t_0)^2 \right.  \\
&& \left. + \frac{1}{3!} j_0 H_0^3 (t-t_0)^3 - \frac{1}{4!} s_0
H_0^4 (t-t_0)^4 \right. \\
&& \left. + \frac{1}{5!} l_0 H_0^5 (t-t_0)^5 + O[(t-t_0)^6]\right\}~, \nonumber
\eea
and its inverse (note the typo in the term $O[(t-t_0)^3]$ in Ref. \cite{ANN06} - should be $j_0/6$ instead of $j_0/3$) reads as
\bea
\label{seriesinva}
&&\frac{a_0}{a(t)} = 1+z = 1 + H_0 (t_0-t) + H_0^2
\left(\frac{q_0}{2} +1 \right) (t_0-t)^2 \nonumber \\
&& + H_0^3 \left(q_0 + \frac{j_0}{6} + 1 \right) (t_0-t)^3 \\
&& + H_0^4 \left(1 + \frac{j_0}{3} + \frac{q_0^2}{4} + \frac{3}{2} q_0
+ \frac{s_0}{24} \right) (t_0-t)^4 \nonumber \\
&& + H_0^5 \left[1 + \frac{l_0}{120} + \frac{s_0}{12} + \frac{j_0}{2}\left(1+\frac{q_0}{3} \right) \right. \nonumber \\
&& \left. + q_0 \left(2 + \frac{3}{4} q_0 \right) \right] (t_0 - t)^5 + O[(t_0-t)^6]~. \nonumber
\eea
The inversion of (\ref{seriesinva}) gives
\bea
\label{tVSz}
&&t_0 - t = \frac{1}{H_0} \left\{ z - \left( 1 + \frac{q_0}{2} \right) z^2 + \left[ q_0 \left(1 + \frac{q_0}{2} \right) + 1 - \frac{j_0}{6} \right] z^3 \right. \nonumber \\
&& \left. + \left[ \frac{5}{12} q_0 j_0 - \frac{s_0}{24} + \frac{j_0}{2} - \frac{5}{8} q_0^3 - \frac{3}{2} q_0 (q_0+1) - 1 \right] z^4 \right. \nonumber \\
&& \left. + \left[ 1 - \frac{l_0}{120} + \frac{s_0}{6} - j_0 \left( 1 - \frac{j_0}{12} \right) - j_0 q_0 \left( \frac{5}{3} + \frac{7}{8} q_0 \right)  \right. \right. \nonumber \\
&& \left. \left. + \frac{1}{8} q_0 s_0 + q_0 \left( 2 + 3 q_0 + \frac{5}{2} q_0^2 + \frac{7}{8} q_0^3 \right) \right]z^5 + O(z^6) \right\}~.
\eea
In the Friedmann universe and also in VSL theory \cite{BM99} the luminosity distance is defined as
\begin{equation}
D_L = (1+z)a_0 r = (1+z)a_0 S_K(\chi)  , \label{lum}
\end{equation}
where $r$ is the radial distance from a source to an observer, and
\bea
S_K(\chi) = \left\{
        \begin{array}{l}
            \frac{1}{\sqrt{K}}\sin(\sqrt{K}\chi),  K=+1\\
            \chi,  K = 0\\
            \frac{1}{\sqrt{|K|}}\sinh(\sqrt{|K|}\chi),  K=-1
        \end{array}\right. \
\eea
>From the null geodesic equation in the Friedmann universe we have
\be
\label{geodeq}
\int_{t_e}^{t_0} \frac{c(t) dt}{a(t)} = \int_{0}^{r}
\frac{dr}{\sqrt{1-Kr^2}} = \chi(r) = S_K^{-1}(r)~,
\ee
so that we can calculate
\be
r = S_K(\chi) = S_K \left( \int_{t_e}^{t_0} \frac{c(t) dt}{a(t)} \right)~,
\ee
which can further be expanded in series for small distances as
\bea
\label{rser}
&&r = \left( \int_{t_e}^{t_0} \frac{c(t) dt}{a(t)}
\right) - \frac{K}{3!} \left( \int_{t_e}^{t_0} \frac{c(t) dt}{a(t)}
\right)^3 \nonumber \\
&& + \frac{K^2}{5!} \left( \int_{t_e}^{t_0} \frac{c(t) dt}{a(t)}
\right)^5 + O \left[ \left( \int_{t_e}^{t_0} \frac{c(t) dt}{a(t)}
\right)^7 \right] \nonumber \\
&& \equiv r_1 + r_3 + r_5 + r_7 + \ldots
\eea
Now, the problem reduces to a calculation of the integral (\ref{geodeq}) in terms of the time flight of a light ray $(t_0 - t)$. For this sake, we use
series expansion (\ref{seriesinva}) with the ansatz (\ref{c(t)}) for the speed of light dependence, i.e.,
\bea
\label{r1}
&&r_1 = \int_{t_e}^{t_0} \frac{c(t) dt}{a(t)} = \frac{c_0}{a_0^n} \int_{t_e}^{t_0} a^{n-1}(t) dt \\
&& = \frac{c_0}{a_0} \int_{t_e}^{t_0} dt
\left\{ 1 + (n-1) H_0 (t-t_0) \right. \nonumber \\
&& \left. - \frac{1}{2} (n-1) H_0^2
\left(q_0 - n + 2 \right) (t-t_0)^2 \right. \nonumber \\
&& \left. + \frac{1}{2} (n-1) H_0^3 \left[\frac{j_0+(n-2)(n-3)}{3} -(n-2)q_0 \right] (t-t_0)^3
\right.  \nonumber \\
&& \left. + \frac{1}{2} (n-1) H_0^4 \left[\frac{(n-2)j_0}{3} + \frac{1}{2} (n-2) q_0 \left(\frac{q_0}{2} - n + 3 \right) \right. \right. \nonumber \\
&& \left. \left. + \frac{-s_0+(n-2)(n-3)(n-4)}{12} \right] (t-t_0)^4
\right. \nonumber \\
&& \left. + \frac{1}{12}(n-1)H_0^5 \left[ \frac{l_0 + (n-2)(n-3)(n-4)(n-5)}{10} \right. \right. \nonumber \\
&& \left. \left. - q_0 (n-2)(n-3)(n-4) + (n-2)(n-3)\left(j_0 + \frac{3}{2} q_0^2 \right) \right. \right. \nonumber \\
&& \left. \left. - (n-2)\left(j_0 q_0 + \frac{s_0}{2} \right) \right](t_0 - t)^5 + O[(t-t_0)^6]\right\}~, \nonumber
\eea
which integrates to give
\bea
\label{r01}
&&r_1 = \frac{c_0}{a_0} \left\{ (t_0 - t) + \frac{1-n}{2} H_0 (t_0 - t)^2 \right. \\
&& \left. + \frac{1-n}{6} H_0^2 \left(q_0 - n + 2 \right) (t_0 - t)^3 \right. \nonumber \\
&& \left. + \frac{1-n}{8} H_0^3 \left[\frac{j_0+(n-2)(n-3)}{3} -(n-2)q_0 \right] (t_0 - t)^4
\right.  \nonumber \\
&& \left. - \frac{1-n}{10} H_0^4 \left[\frac{(n-2)j_0}{3} + \frac{n-2}{2} q_0 \left(\frac{q_0}{2} - n + 3 \right) \right. \right. \nonumber \\
&& \left. \left. + \frac{-s_0+(n-2)(n-3)(n-4)}{12} \right] (t_0 - t)^5
\right. \nonumber \\
&& \left. - \frac{1}{72}(n-1)H_0^5 \left[ \frac{l_0 + (n-2)(n-3)(n-4)(n-5)}{10} \right. \right. \nonumber \\
&& \left. \left. - q_0 (n-2)(n-3)(n-4) + (n-2)(n-3)\left(j_0 + \frac{3}{2} q_0^2 \right) \right. \right. \nonumber \\
&& \left. \left. - (n-2)\left(j_0 q_0 + \frac{s_0}{2} \right) \right](t_0 - t)^6 + O[(t-t_0)^7]\right\}~,\nonumber
\eea
Notice that using (\ref{tVSz}) and (\ref{r01}) gives
\bea
\label{r03}
&& r_3 = - \frac{K}{3!} \left( \int_{t_e}^{t_0} \frac{c(t) dt}{a(t)} \right)^3 = - \frac{K}{6}  \frac{c_0^3}{a_0^3}(t_0 - t)^3 \nonumber \\
&& \times \left\{ \frac{3H_0^2(1-n)}{2} \left[ \frac{1}{2} (1-n) + \frac{1}{3} (q_0 - n + 2) \right] (t_0 - t)^2  \right. \nonumber \\
&& \left. + \frac{3H_0(1-n)}{2} (t_0 - t) + 1 \right\} + O[(t-t_0)^7\\
&&= \frac{\Omega_{K0}}{2} \frac{c_0}{a_0 H_0} \left\{ - \frac{1}{3} z^3 + \frac{1}{2}  \left(q_0 + n + 1\right) z^4 \right. \nonumber \\
&& \left.- \underline{\frac{1}{12}} \left[ 7 - 2j_0 + n(12 + 5n +10q_0) + q_0(14 + 9q_0) \right] z^5 \right. \nonumber \\
&& \left. +\underline{\frac{1}{24}} \left[ 15 + s_0 - j_0(14q_0 +13 + 7n) + q_0(45 +58q_0 + 28q_0^2) \right. \right. \nonumber \\
&& \left. \left. + n(34 + 25n + 6n^2) + n q_0(57 + 18n + 32q_0) \right] z^6 + O(z^7) \right\} , \nonumber
\eea
and
\bea
\label{r05}
&& r_5 = \frac{K^2}{5!} \left( \int_{t_e}^{t_0} \frac{c(t) dt}{a(t)} \right)^5 \\
&& = \frac{\Omega_{K0}^2}{24} \frac{c_0}{a_0 H_0} \left[ \frac{1}{5} z^5 - \frac{1}{2} \left(q_0 + n + 1\right) z^6 + O(z^7) \right]. \nonumber
\eea
The physical distance $D$ which is travelled by a light ray emitted at time $t_e$, and received at time $t_0$ in the VSL theory is given by
\bea
\label{D}
D &=& \int_{t_e}^{t_0} c(t) dt = c_0 \left\{ (t_0 - t_e) - \frac{n}{2} H_0 (t_0 - t_e)^2 \right. \nonumber \\
&-& \left. \frac{n}{6} H_0^2 (q_0 - n + 1)(t_0 - t_e)^3 \right. \nonumber \\
&+& \left. \frac{n}{8} H_0^3 \left[ \frac{j_0 + (n-1)(n-2)}{3} - (n-1) q_0 \right] (t_0 - t_e)^4 \right. \nonumber \\
&-& \left. \frac{n}{10} H_0^4 \left[ \frac{(n-1)j_0}{3} + \frac{n-1}{2} q_0 \left(\frac{q_0}{2} - n + 2 \right) \right. \right. \nonumber \\
&+& \left. \left. \frac{-s_0+(n-1)(n-2)(n-3)}{12} \right] (t_0 - t)^5 \right. \nonumber \\
&+& \left. O[(t_0-t_e)^6] \right\}~.
\eea
The radial distance (\ref{rser}) as a function of redshift $z$ (cf. (\ref{tVSz})) reads as
\bea
\label{rz}
&& r(z) =  \frac{c_0}{a_0H_0} \times \left\{ z - \frac{1}{2} (q_0 + n + 1) z^2 \right. \nonumber \\
&& \left. + \left[ \frac{q_0^2}{2} + \frac{1}{3} (2q_0 + 1) - \frac{j_0}{6}
- \frac{\Omega_{K0}}{6} + \frac{n}{6} (2q_0 + n + 3) \right] z^3 \right. \nonumber \\
&& \left. + \left[ \frac{5}{12} q_0 j_0 - \frac{s_0}{24} + \frac{3}{8} j_0
- \frac{5}{8} q_0^3 - \frac{3}{4} \left( \frac{3}{2} q_0^2 + q_0 +
\frac{1}{3} \right) \right. \right. \nonumber \\
&& \left. \left. + \frac{1}{4} \left( q_0 + n + 1 \right) \Omega_{K0}
+ n \left( \frac{j_0}{8} - \frac{3}{8} q_0^2 - \frac{5}{8} q_0 - \frac{11}{24} \right) \right. \right. \nonumber \\
&& \left. \left. - \frac{n^2}{4} \left( \frac{q_0}{2} + 1 + \frac{n}{6} \right) \right] z^4
+ \left[ \frac{1}{5} - \frac{l_0}{120} + s_0 \left( \frac{2}{15} + \frac{1}{30} n + \frac{1}{8} q_0 \right) \right. \right. \nonumber \\
&& \left. \left. - \frac{1}{3} j_0 q_0 \left(4 + n + \frac{21}{8} q_0 \right) + j_0 \left(\frac{j_0}{12} - \frac{3}{5} - \frac{7}{20} n - \frac{1}{20} n^2 \right) \right. \right. \nonumber \\
&& \left. \left. + q_0 \left( \frac{4}{5} + \frac{9}{5} q_0 + 2q_0^2 + \frac{7}{8} q_0^3 \right)  \right. \right.  \\
&& \left. \left. + n q_0 \left( \frac{13}{15} + \frac{3}{10} n + \frac{1}{30} n^2 + \frac{21}{20} q_0 + \frac{3}{20} n q_0 + \frac{1}{2} q_0^2 \right) \right. \right. \nonumber \\
&& \left. \left. - \frac{1}{24} \Omega_{K0} \left( 7 - 2j_0 + n(12 + 5n +10q_0) + q_0(14 + 9q_0) \right) \right. \right. \nonumber \\
&& \left. \left. + n \left( \frac{5}{12} + \frac{7}{24} n + \frac{1}{12} n^2 + \frac{1}{120} n^3 \right) + \frac{1}{120} \Omega_{K0}^2 \right] z^5 + O(z^6) \right\}~. \nonumber
\eea
Finally, from (\ref{lum}) and (\ref{rz}), we obtain the
fifth-order in redshift $z$ formula for the luminosity distance as
\bea
\label{lumgeneralz4}
&&D_{L}(z) = c_0 \frac{z}{H_0} \times \left\{1 + \frac{1}{2} (1- q_0 - n)z \right. \nonumber \\
&&+ \left. \frac{1}{6} \left[q_0(3q_0 + 2n + 1) + (n^2 - j_0 - 1) - \Omega_{K0} \right]z^2
\right. \nonumber \\
&& \left. + \frac{1}{24} \left[ 5j_0(2q_0 + 1) - s_0 - 15q_0^2 (q_0 + 1) + 2(1 - q_0)
\right. \right. \nonumber \\
&& \left. \left. + 2 \Omega_{K0}(3q_0 + 3n + 1) + n (3 j_0 - 9 q_0^2 - 7 q_0 + 1 ) \right. \right. \nonumber \\
&& \left. \left. - n^2 (3q_0 + n + 2) \right] z^3 + \frac{1}{120} \left[ - 6 - l_0 + \Omega_{K0}^2 \right. \right. \nonumber \\
&& \left. \left. - \Omega_{K0} (5 - 10j_0 +30n +25n^2 + 40 q_0 + 50 n q_0 + 45 q_0^2) \right. \right. \nonumber \\
&& \left. \left. + s_0 (11 + 4n + 15q_0) - j_0 (27 - 10 j_0 + 27 n + 6 n^2 ) \right. \right. \nonumber \\
&& \left. \left. + 5 j_0 q_0 ( 22 - 8n - 21q_0 ) + n (-5 + 5n + 5n^2 + n^3 ) \right. \right. \nonumber \\
&& \left. \left. + n q_0 (29 + 21n + 4n^2 + 81 q_0 + 18 n q_0 + 60 q_0^2 ) \right. \right. \nonumber \\
&& \left. \left. + q_0 (6 + 81 q_0 + 165q_0^2 + 105 q_0^3 ) \right]z^4 + O(z^5) \right\}~.
\eea
The formula (\ref{lumgeneralz4}) for $n=0$ agrees with the third-order in $z$ formula (37) of
\cite{jerk}, with the fourth-order in $z$ formula (7) of \cite{snap}, and the fifth-order in $z$ formula (A6) of \cite{salvatore}.
It also agrees with the formula (39) of \cite{AJIII} for
$\Omega_{st0} = \Omega_{g0} = \Omega_{ph0} = \Omega_{sp0} = \Omega_{bp0} = \Omega_{hp0}$.
As for the VSL theory it generalizes the formula (7) of Ref. \cite{BM99} to include jerk, snap and lerk (crack).

The series expansion (\ref{lumgeneralz4}) for the luminosity distance $D_L(z)$ can alternatively be obtained by using the method of Kristian and Sachs
\cite{KS,AJIII}. It is worth mentioning that this derivation (starting from formula (\ref{r1})) would have been different if either the ans\"atze $c(t) = c_0 [a(t)]^n$ or $c(t) = c_0 [a(t)]^{n(t)}$ of Ref. \cite{BM99} were applied. In particular, the ansatz $c(t) = c_0 [a(t)]^n$ would not make the parameter $n$ dimensionless and it would enter all our formulas beginning from (\ref{r1}) already in the first order which is not acceptable physically since the physical distance should have the proper unit of length which can be seen from the formula (\ref{D}) for example. A different ansatz for $c(t) = \dot{a}(t)$ was used in Ref. \cite{buchalter}, but it leads to totally different (though exact) expression for the luminosity distance.

It is also interesting to note that the redshift drift formula for the VSL theories \cite{driftVSL} is independent of the choice of the ansatz either in the form of (\ref{c(t)}) or that one given by $c(t) = c_0 [a(t)]^n$ as in Ref. \cite{BM99}.

\section{Discussion}
\setcounter{equation}{0}
\label{discussion}

We have derived a luminosity distance formula (\ref{lumgeneralz4}) for the varying speed of light cosmology. We have used the specific ansatz for the variability of $c(t) = c_0 \left( \frac{a(t)}{a_0} \right)^n$ in order to discuss the effect of varying $c$ onto the redshift change over the evolution of the universe. This formula involves higher order characteristics of expansion such as jerk, snap, and lerk. The second-order luminosity distance in redshift $z$ formula is modified in the way that negative $n$ leads to stronger acceleration of the universe, positive $n$ leads to its stronger deceleration, while in the third (jerk), fourth (snap), and fifth (lerk) order this influence is more complex and includes $n^2$, $n^3$, $n^4$ terms.

Bearing in mind the ansatz (\ref{c(t)}) one can say that the value of the speed of light was different in the past radiation epoch, and then it was gradually reaching its present value $c_0$. As for the parameter $n$ we know that $\mid n \mid  \sim 10^{-5} < 0$ and this value which is compatible with the current observational constraints on $c \propto {\alpha}^{-1}$ \cite{murphy2007,king2012} is rather small. Its influence onto the luminosity distance formula (\ref{lumgeneralz4}) does not seem to be easily detectable in analogy to the redshift drift formula \cite{driftVSL} which requires the minimum value of $\mid n \mid > 0.045$ in order for these VSL models to be distinguished from the $\Lambda$CDM models.

Using both the luminosity distance formula (\ref{lumgeneralz4}) and the relations between observational parameters $\Omega_{m0}$, $\Omega_{\Lambda 0}$, $\Omega_{K0}$, $H_0$, $q_0$, $j_0$, $s_0$, and $n$ (\ref{0th})-(\ref{4th}) which come from the field equations we may conclude the following.

The variability of the speed of light enters in the second order of the expansion of $D_L(z)$ and then appears in every higher-order term. Further, due to the relation (\ref{0th}) which comes directly as a consequence of the Friedmann equation, one can calculate the curvature density parameter $\Omega_{K0}$ provided one knows the values of the matter density parameter $\Omega_{m0}$ and the cosmological constant density parameter $\Omega_{\Lambda 0}$ from other measurements (e.g. $H(z)$ cf. Ref. \cite{farooq}). This would also allow to calculate the VSL parameter $n$ by using the relation (\ref{1st}) as well as the second-order term of $D_L(z)$ expansion (\ref{lumgeneralz4}) (2 equations for 2 unknowns). Having $n$, one then would be able to isolate $H_0$ uniquely. If there was a trouble measuring $\Omega_{\Lambda 0}$, then one would not stick to only measuring the cubic term $O(z^3)$ in the luminosity distance which involves jerk parameter $j_0$, but also to quartic term $O(z^4)$ with snap parameter $s_0$ since both of them include curvature, but the two relations (\ref{2nd}) and (\ref{3rd}) which come from the field eqautions, would help to resolve the problem. Finally, in a hypothetic case mentioned in Ref. \cite{snap} where the curvature term $\Omega_{K0}$ would be mixed with an exotic fluid of cosmic strings (cf. Ref. \cite{AJIII}), one would have to even go to a higher order quintic term $O(z^5)$ in the luminosity distance which involves lerk parameter $l_0$ in order to isolate the curvature of the universe.

All these measurements are the matter of the future investigations of the higher order characteristics of the expansion of the universe known as cosmography \cite{ruth}.

\section{Acknowledgements}

This project was financed by the National Science Center Grant DEC-2012/06/A/ST2/00395. We thank Robert Caldwell for discussions. 



\end{document}